# Learning native eddy-resolving ocean dynamics outperforms coarse emulation and downscaling

Authors

Zilu Liu[1], Tao Zhang[1*], Hailong Liu[1*], Kai Xu[1], Qinghui Lin[2], Maoxue Yu[1], Jiangfeng Yu[1], Jingwei Xie[1], Zhao Jing[1,3], Minghua Zhang[4], Lixin Wu[1,3]

1. Laoshan Laboratory, Qingdao, China
2. Gosci Technology Group, Qingdao, China
3. Frontiers Science Center for Deep Ocean Multispheres and Earth System and Key Laboratory of Physical Oceanography,Ocean University of China, Qingdao, China
4. School of Marine and Atmospheric Sciences, Stony Brook University, Stony Brook, NY, USA

Corresponding authors: Tao Zhang (tzhang@lsnl.cn), Hailong Liu (hliu@lsnl.cn)

## Abstract

Eddy-resolving ocean simulations are essential for representing mesoscale dynamics that regulate ocean circulation, energy transport, and climate variability. However, their scale poses challenges for generation, storage, and accessibility. Data-driven artificial intelligence (AI) methods offer a pathway to address these limitations, but it remains unclear whether high-resolution ocean fields can be faithfully represented by evolving ocean states at coarse resolution followed by spatial downscaling, or whether dynamical evolution needs to be learned directly from native high-resolution simulations. We show that directly learning dynamical evolution on a 0.1° grid provides a more faithful representation of mesoscale variability than evolving ocean states on a 1° grid followed by downscaling reconstruction. We developed a full-depth 1° emulator and a native 0.1° sea surface temperature (SST) emulator, both of which remained stable over long-term free-running simulations without significant drift, providing reliable baselines for this comparison. Direct learning at 0.1° preserves mesoscale energy, spectral characteristics, and frontal gradients, whereas the coarse-resolution evolution followed by downscaling approach recovers large-scale patterns but fails to reproduce key mesoscale signatures. We show that the learned mesoscale statistics remain consistent across diverse hardware platforms, supporting the reproducibility and portability of AI-based ocean simulations. These results suggest that downscaling of low-resolution large-scale states alone is insufficient to represent ocean mesoscale variability, supporting the need to learn dynamical evolution directly in a high-resolution state space. These findings highlight the continued importance of high-resolution physics ocean models as a foundation for AI-based ocean simulation, providing dynamically resolved training data for physically faithful high-resolution emulators.

## INTRODUCTION

Reliable and reusable high-resolution ocean fields are fundamental to studies of ocean dynamics, circulation-driven biogeochemical cycles, and ocean-state prediction and forecasting. Mesoscale eddies, fronts, and filaments strongly influence the transport and redistribution of heat, salt, nutrients, and other tracers, and contribute to the regulation of air-sea exchange *(1–4)*. Explicit representation of these processes generally requires horizontal grid spacing of O(10 km) or finer, although the exact threshold depends on latitude, the local Rossby radius of deformation, and the effective resolution of the model *(5, 6)*. At such resolutions, computational, storage, and data-transfer demands escalate rapidly, limiting the generation, dissemination, and repeated use of global eddy-resolving simulations *(7)*.

Data-driven artificial intelligence (AI) provides a new route for ocean simulation. Rather than explicitly solving the full set of discretized governing equations, ocean emulators learn the temporal evolution of ocean states from physics model simulations or reanalysis data and then advance the system autoregressively at a small fraction of the cost of conventional integration *(8, 9)*. Recent models, such as Samudra, ORCA-DL, and Samudra 2, have reproduced large-scale circulation and major modes of climate variability *(8–10)*. Their stability over multiyear to decadal, or even longer integrations highlights their potential as efficient complements to traditional numerical ocean models. Their significance may also extend beyond computational acceleration. A fixed-weight emulator can serve as a distributable and repeatedly executable representation of simulation information, allowing users to regenerate trajectories from prescribed initial states and external forcing and thereby changing how expensive numerical simulations are stored, distributed, and reproduced *(11, 12)*. This possibility, however, depends on the statistical and physical reliability of the generated ocean fields and on the reproducibility of inference across computing platforms.

Current high-resolution AI ocean research has developed mainly toward two objectives: state forecasting and model emulation. High-resolution AI ocean forecast systems, including WenHai at 1/12° and FuXi-Ocean, are designed to improve forecasts initialized from analyzed ocean states, and they are reinitialized from high-resolution analysis fields for each forecast cycle *(13, 14)*. In this setting, part of the mesoscale structure in a forecast can be inherited directly from the high-resolution initial condition and then propagated over a finite lead time *(13)*. The appearance of well-defined eddies and fronts in short- or medium-range forecasts therefore does not by itself demonstrate that a model can autonomously generate and maintain mesoscale variability without repeated reinitialization. Answering that question requires free-running simulations in which both climate-state stability and mesoscale statistics are evaluated *(8, 13)*.

Ocean-model emulators, by contrast, are intended to learn the dynamical evolution encoded by numerical ocean models and to retain core scientific-computing capabilities, including scenario simulation, sensitivity analysis, parameter-space exploration, long-term integration, and coupling to other physical components *(8, 15)*. Global ocean emulators have made substantial progress toward this goal. For example, the 1° configuration of Samudra and ORCA-DL can simulate large-scale global ocean states, but their grids are too coarse to explicitly resolve eddy-resolving mesoscale structures *(5, 8, 9)*. One of the closest global attempts is Samudra 2 at 1/4°. This resolution is generally considered eddy-permitting rather than eddy-resolving, and existing evaluations have focused mainly on the mean state, long-term stability, or larger-scale variability rather than systematically testing the fidelity of oceanic mesoscale variability represented by the eddy kinetic energy, spectral structure, and frontal gradients *(5, 10)*. It therefore remains unknown whether an AI ocean model can autonomously generate and sustain ocean mesoscale structures with eddy-resolving statistical fidelity during long free-running simulations.

Two pathways can be used within the broader strategy of obtaining eddy-resolving AI ocean simulations (Fig. 1). Pathway 1, native learning (NL), trains an emulator directly on eddy-resolving numerical-model data, such that both the learned dynamical evolution and the final output reside on a high-resolution grid *(10, 16)*. Pathway 2, downscaling (DS), first trains an emulator on low-resolution data and then uses a separate deterministic or generative downscaling module to map each coarse state onto an eddy-resolving grid. Related coarse-to-fine approaches are increasingly used in Earth-system modeling. NVIDIA Earth-2, for example, includes CorrDiff for regional weather downscaling from ~25km global fields to ~2km fields, and cBottle for global climate generation, which uses a cascaded architecture to generate ~100km atmospheric states and downscale them to ~5 km. *(12, 17)*. The pathways differ fundamentally in how fine-scale structures are generated: Pathway 1 evolves eddy-resolving states, allowing mesoscale structures to emerge as part of the learned dynamics, whereas Pathway 2 evolves only coarse-resolution states and reconstructs unresolved structure afterward. Although both pathways can produce outputs on the same high-resolution grid, it remains unclear whether mesoscale fidelity is controlled primarily by the final output resolution or by the resolution at which dynamical evolution is learned. A visually detailed high-resolution field may not reproduce the correct mesoscale energy, high-wavenumber spectrum, frontal intensity, or temporally consistent evolution *(13, 17)*. Distinguishing these possibilities motivates the controlled comparison presented here.

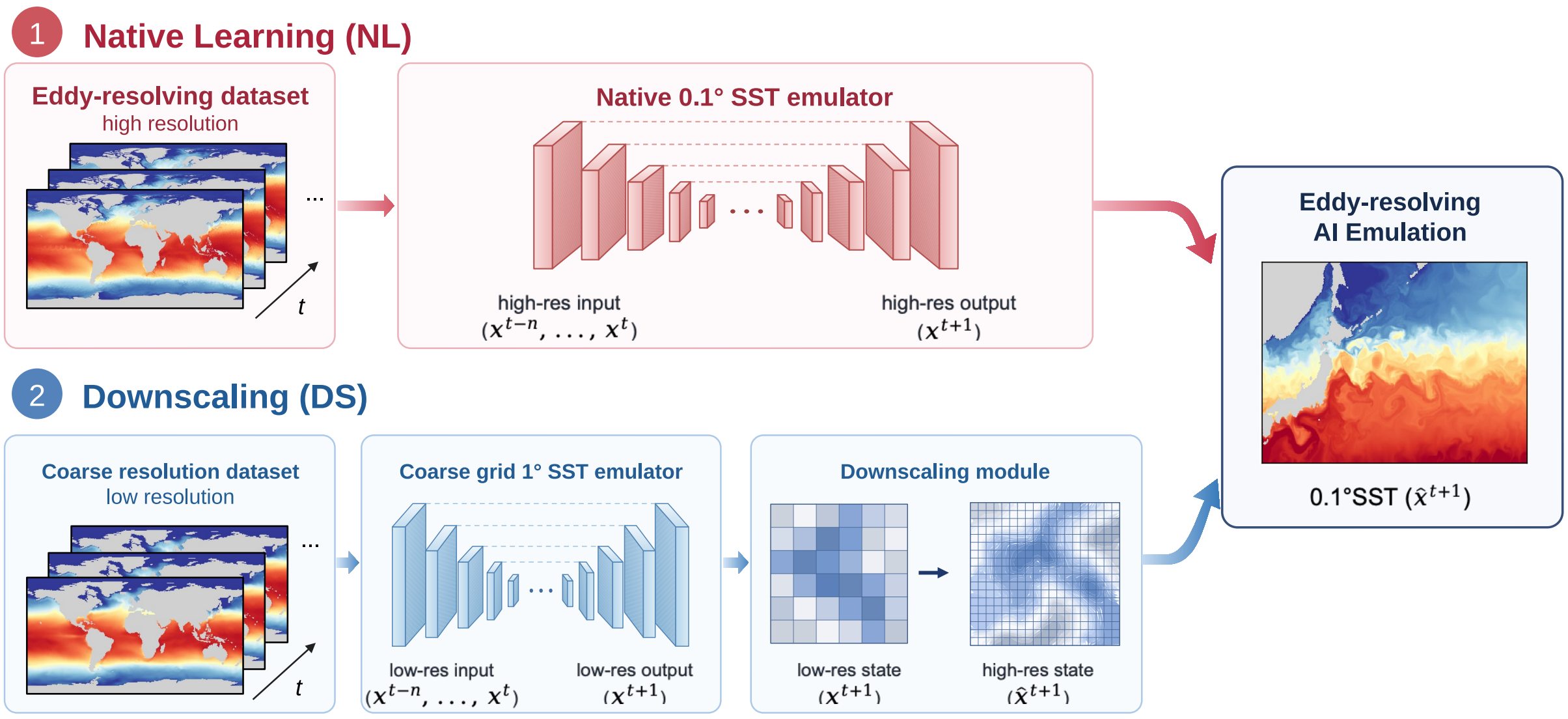

**Fig. 1. Two strategies for eddy-resolving AI ocean simulation.** Pathway 1, native learning (NL), trains an AI emulator directly on eddy-resolving data, such that both the learned temporal evolution and the output are high resolution. Pathway 2, downscaling (DS), first learns temporal evolution from low-resolution data and then maps each low-resolution state to an eddy-resolving grid using a separate downscaling module. Both pathways target eddy-resolving output, but the resolution at which the dynamics are learned differs.

Here, we conduct a controlled comparison of these two pathways to determine how the resolution at which dynamics are learned affects the representation of mesoscale structures. We construct a full-depth 1° emulator and a native 0.1° SST emulator trained directly on the corresponding ocean simulations. Both remain stable in long free-running simulations without significant climate drift, providing credible large-scale baselines for mesoscale analysis. Within a unified diagnostic framework, we evaluate mesoscale energy, wavenumber spectra, and frontal gradients, and further test the statistical reproducibility of a fixed-weight model across different hardware and software configurations. We show that directly learning dynamical evolution on the 0.1° grid provides higher overall mesoscale fidelity than evolution on the 1° grid followed by downscaling reconstruction. Although downscaling recovers large-scale spatial morphology, it substantially underestimates mesoscale energy, high-wavenumber spectral power, and strong frontal gradients. These results suggest that post hoc downscaling of low-resolution large-scale states alone is insufficient to fully represent mesoscale variability, highlighting the importance of learning dynamical evolution directly in a high-resolution state space.

## RESULTS

### Stable large-scale climate baselines at 1° and 0.1° resolutions

We construct baseline emulators trained directly on 1° and 0.1° swLICOM ocean simulations *(18, 19)*. The full-depth 1° emulator is an ocean emulator trained on monthly mean data, representing five ocean variables across 30 depth levels, whereas the native 0.1° SST emulator is an SST-only emulator trained on daily mean data. We first examined whether the emulators established stable large-scale climate baselines for the subsequent comparison. Fig. 2 summarizes their global-mean sea surface temperature during the test periods and during 100-year free-running simulations initialized from the test state. This evaluation tests both the ability to reproduce the temporal evolution of the reference simulations and the stability of long autoregressive integrations.

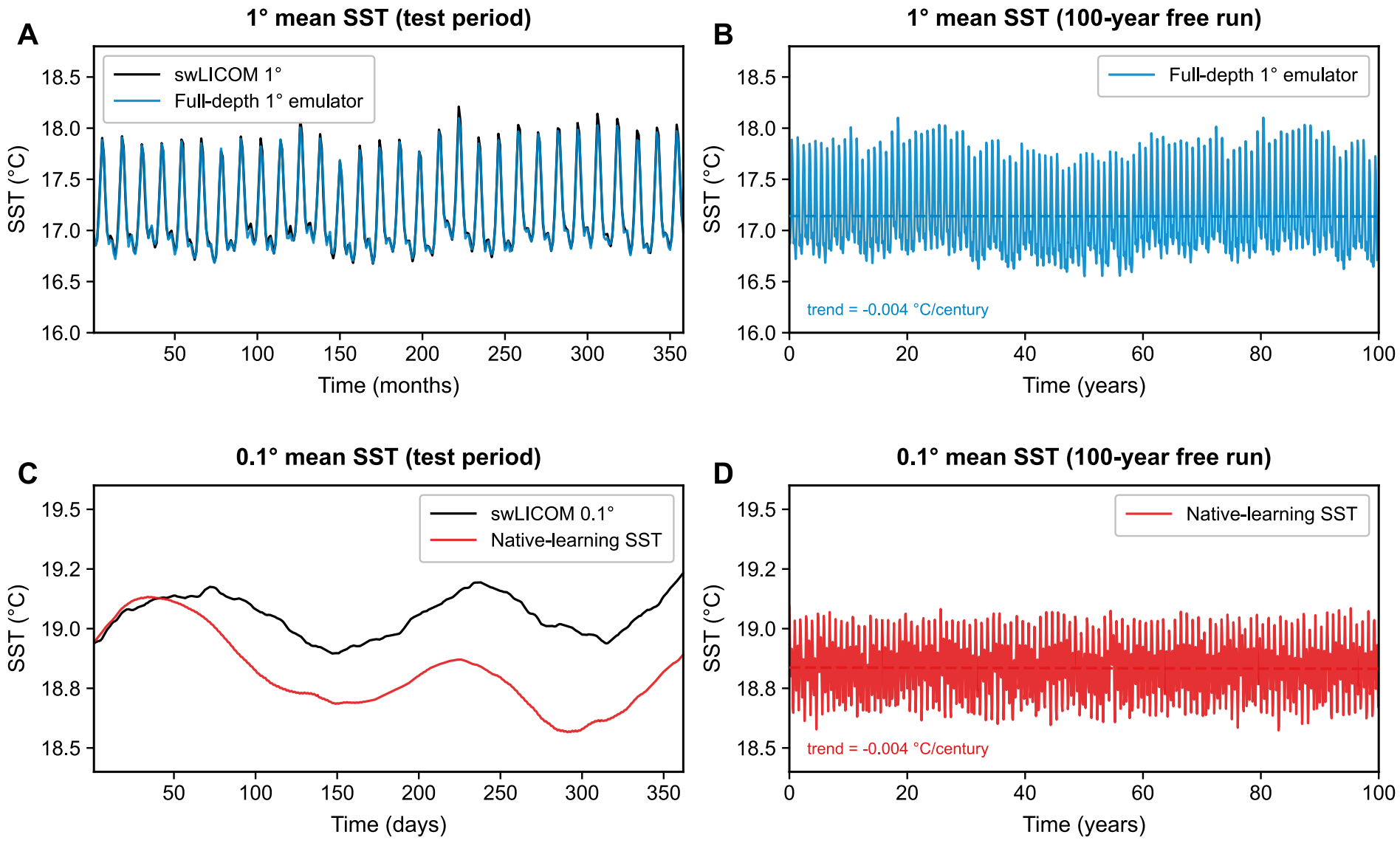

**Fig. 2. Test-period performance and centennial free-running stability of the full-depth 1° emulator and native 0.1° SST emulator. (A)** Global-mean SST over years 95–124, the first 30 years of the 1° 100-year rollout, for the swLICOM 1° simulation (black) and the full-depth 1° emulator inference (blue). **(B)** Global-mean SST over the full 100-year 1° rollout (years 95–194), with a fitted emulator trend of −0.004 °C century$^{-1}$; the black series denotes the swLICOM 1° reference. **(C)** Global-mean SST during the independent 2023 test period for the swLICOM 0.1° simulation (black) and Native-learning SST (red). **(D)** Global-mean SST during the 100-year free run of the native 0.1° SST emulator, with a fitted trend of −0.004 °C century$^{-1}$; the black series denotes the swLICOM 0.1° reference. Both free-running simulations retain bounded seasonal variability without substantial long-term drift.

During the test periods, both emulators were consistent with the main temporal variations of their corresponding swLICOM simulations (Fig. 2, A and C). The full-depth 1° emulator nearly reproduced the reference sequence, including the seasonal cycle and its amplitude modulation. The global-mean SST of the native 0.1° SST emulator was systematically lower than that of swLICOM 0.1°, indicating a mean bias, but it retained similar low-frequency variations and major turning points. Thus, despite the substantially larger state space at 0.1°, the emulator preserved the large-scale temporal evolution of the reference simulation. The spatial distribution of the mean-state bias, including the equatorial Pacific cold bias at 0.1°, is shown in Fig. S1.

The long free runs further examined whether accumulated errors destabilized the simulated climate. Over 100 years, both the full-depth 1° emulator and native 0.1° SST emulator retained bounded seasonal cycles without persistent warming, cooling, or growth in seasonal amplitude (Fig. 2, B and D). The fitted linear trend in global-mean SST was −0.004 °C century$^{-1}$ at both resolutions, which is small relative to the seasonal variability and indicates negligible systematic drift over the integration.

These results establish stable large-scale climate baselines for the full-depth 1° emulator and the native 0.1° SST emulator. For the latter, this stability provides context for the mesoscale diagnostics below, although stability of global-mean SST alone does not demonstrate fidelity of mesoscale structures.

### The native 0.1° SST emulator sustains mesoscale structures

A stable global-mean climate state is necessary for credible simulation but does not by itself establish fidelity of mesoscale eddies, fronts, and filaments *(5, 6)*. We therefore moved from global integral measures to local spatial structure and mesoscale statistics, asking whether the native 0.1° SST emulator continued to produce eddy-resolving spatial variability during free-running integration.

Fig. 3 compares SST snapshots in the Kuroshio Extension. At 1°, both swLICOM and the emulator represented the current system primarily as a broad and smooth meridional temperature transition, capturing the large-scale position and width of the frontal zone (Fig. 3A). At 0.1° resolution, swLICOM contained pronounced frontal meanders, rolled-up eddies, and filamentary structures. The native 0.1° SST emulator produced a similarly organized field and maintained abundant small-scale temperature variability on both sides of the front (Fig. 3B). The high-resolution emulator therefore did not merely preserve a smooth large-scale background; it continued to express clear mesoscale morphology during the free run. Regional snapshots extend this comparison to other current systems and the South Pacific subtropical gyre (Fig. S2).

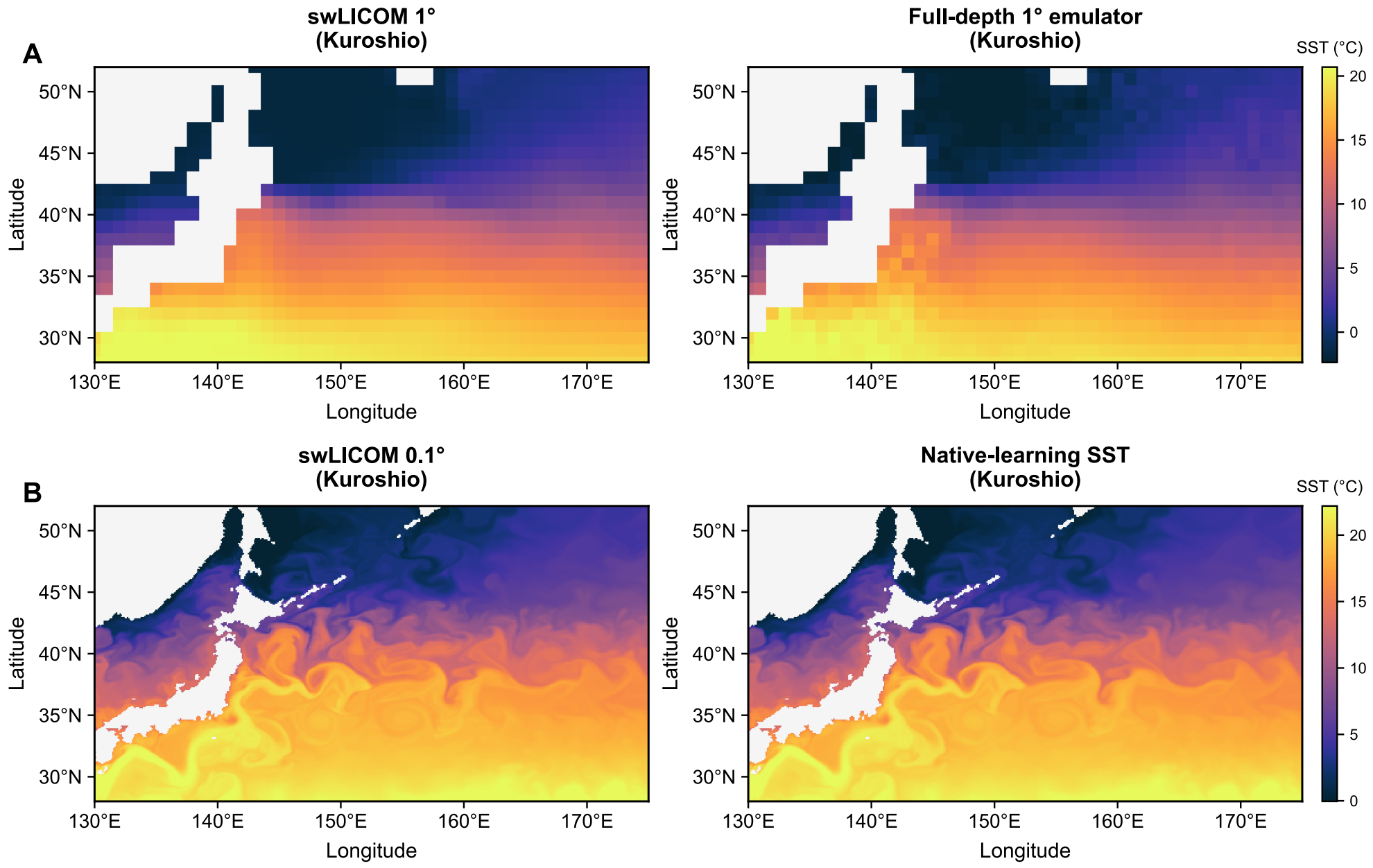

**Fig. 3. SST structure in the Kuroshio Extension at different resolutions. (A)** SST snapshots from the swLICOM 1° simulation (left) and the full-depth 1° emulator inference (right). **(B)** SST snapshots from the swLICOM 0.1° simulation (left) and Native-learning SST (right). The 1° fields contain broad, smooth frontal gradients, whereas the native 0.1° SST emulator reproduces the eddies, frontal meanders, and filaments present in the eddy-resolving reference. White areas denote land.

Spatial-pattern similarity alone is insufficient to establish statistical fidelity. We therefore quantified the native 0.1° SST emulator using three complementary diagnostics (Fig. 4): SST power spectra to characterize the distribution of variance across spatial scales, the spatial probability density function (PDF) of time-mean SST-gradient magnitude to characterize persistently strong frontal gradients, and energy integrated over wavelengths of 50–200 km to quantify the overall strength of mesoscale variability *(1, 20, 21)*. We define $G(x) = \mathrm{mean}_t(|\nabla \mathrm{SST}(x, t)|)$, where the mean is taken over the evaluation period at each ocean grid point. The gradient PDF describes the spatial distribution of $G$ within each analysis display box.

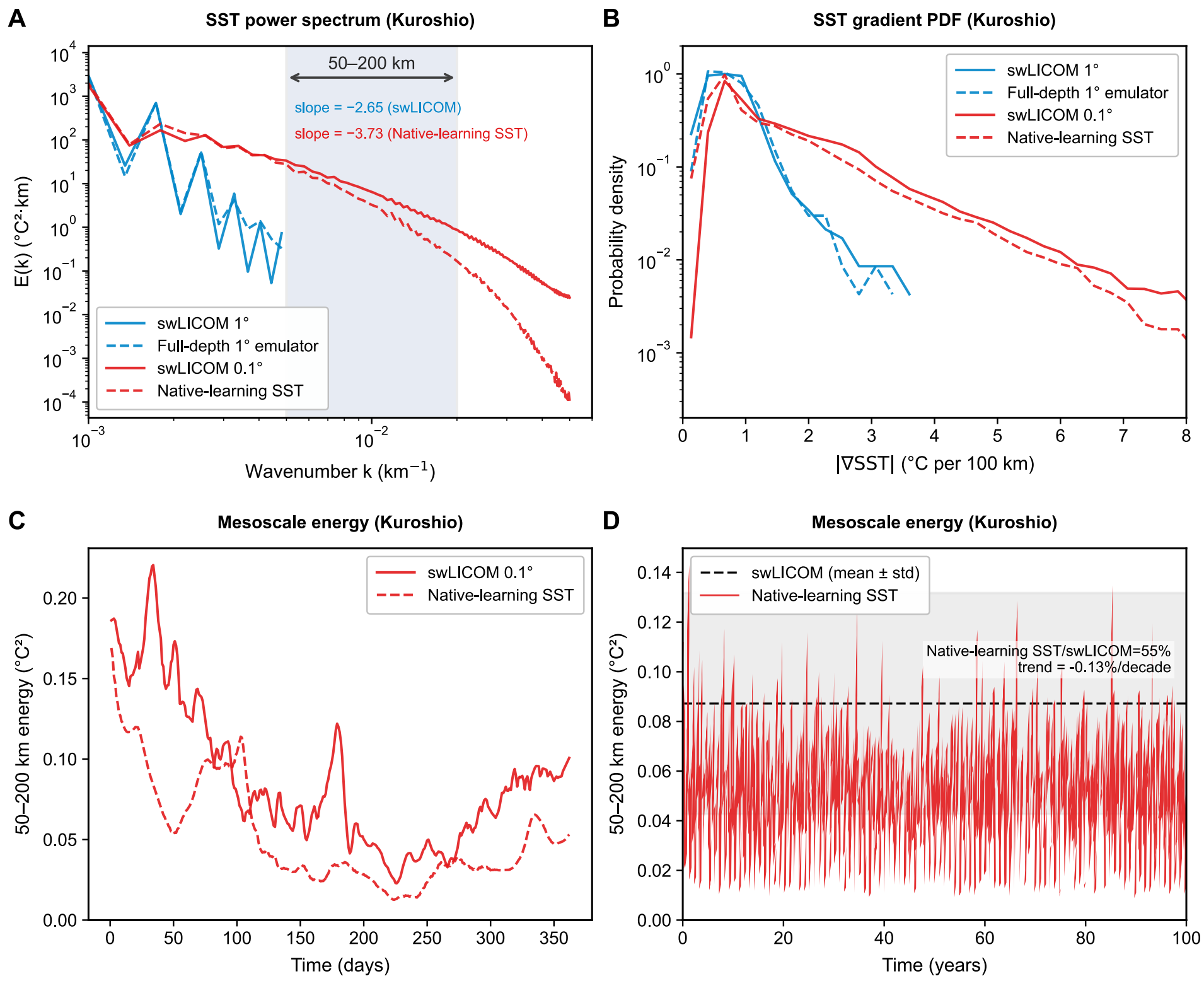

**Fig. 4. Retention of Kuroshio Extension mesoscale statistics by the native 0.1° SST emulator.** Blue and red denote the 1° and 0.1° fields, respectively; solid lines denote swLICOM simulations and dashed lines denote full-depth 1° emulator output and Native-learning SST at the respective resolutions. **(A)** SST power spectra; the gray band marks wavelengths of 50–200 km, and the fitted slopes for the swLICOM 0.1° simulation and Native-learning SST are −2.65 and −3.73, respectively. **(B)** Spatial probability densities of time-mean SST-gradient magnitude G within 130–175° E, 28–52° N. Means cover the respective test periods: model years 95–124 at 1° and 2023 at 0.1°. The 1° curves use 30 bins over 0–8 °C per 100 km; the 0.1° curves use 45 bins over 0–12 °C per 100 km. The displayed range is 0–8 °C per 100 km, not the maximum gradient. **(C)** Time evolution of 50–200 km mesoscale energy during the test period. **(D)** Mesoscale energy during the 100-year free run of the native 0.1° SST emulator; the red curve denotes the emulator inference, while the black dashed line and gray shading denote the swLICOM mean and ±1 standard deviation, respectively.

The native 0.1° SST emulator retained a continuous high-wavenumber tail over a broad range and was broadly consistent with the swLICOM 0.1° spectrum at low and intermediate wavenumbers (Fig. 4A). Within the 50–200 km band, the fitted spectral slopes were −2.65 for swLICOM 0.1° and −3.73 for the emulator. The steeper spectral decay in the emulator indicates a systematic loss of variance toward smaller scales. By comparison, the 1° fields covered only a limited portion of this wavelength band and could not represent its full mesoscale spectral structure *(5, 6)*.

The spatial distributions of time-mean SST-gradient magnitude showed the same tendency (Fig. 4B). The native 0.1° SST emulator reproduced a long-tailed distribution, indicating sustained frontal structure during autoregressive integration. However, the emulator had lower probability density than swLICOM 0.1° over the displayed stronger-gradient range. Persistently strong frontal gradients were therefore spatially underrepresented, consistent with the steeper high-wavenumber spectrum.

The 50–200 km band energy further tested whether these structures persisted through time. During the test period, the emulator followed the major phases of increasing and decreasing mesoscale energy in the reference simulation, but its amplitude was systematically lower over most of the period (Fig. 4C). In the 100-year free run, the band energy remained bounded and neither collapsed to zero nor increased without limit (Fig. 4D). Its long-term mean was approximately 55% of the swLICOM reference, with a fitted trend of −0.13% decade$^{-1}$.

In the Kuroshio Extension, the snapshots, spectra, gradient distributions, and long-term band energy show that the native 0.1° SST emulator can autonomously generate and sustain mesoscale structures rather than merely inheriting high-resolution detail near initialization. At the same time, the steeper spectral slope, weaker gradient tail, and approximately 45% energy deficit reveal that the emulator remains overly smooth and does not fully recover the intensity of mesoscale activity in the eddy-resolving reference. Regional diagnostics show that the degree of variance retention varies among ocean regimes (Fig. S3).

**Native learning preserves mesoscale structures better than downscaling**

Having established that the native 0.1° SST emulator generates and maintains mesoscale variability, we next compared its mesoscale fidelity with that of the downscaling pathway on the same high-resolution output grid. We compared the native learning and downscaling pathways introduced above, with their respective outputs termed Native-learning SST and Downscaled SST.

The two pathways were aligned in the key experimental conditions. Both produced final outputs on the 0.1° grid, their training data originated from the same segment of the swLICOM simulation, and they were evaluated over the same period with identical mesoscale diagnostics. The downscaling module was itself trained on paired 1°/0.1° samples and therefore had access to high-resolution targets from the same source. This design compares the two prediction pathways on a common output grid, with both pathways having access to high-resolution training information from the same source.

The spatial snapshots first revealed a clear difference between the pathways (Fig. 5A). Downscaled SST recovered the large-scale position of the Kuroshio Extension, the meridional temperature distribution, and the approximate width of the frontal zone, demonstrating that the downscaling module reconstructed the major spatial morphology from the 1° output. Its field was nevertheless substantially smoother than both swLICOM 0.1° and Native-learning SST, with fewer continuous filaments, closed eddies, and sharp local temperature gradients along the front. Global variability and time-mean gradient maps provide complementary spatial evidence for the contrast between the pathways (Fig. S4).

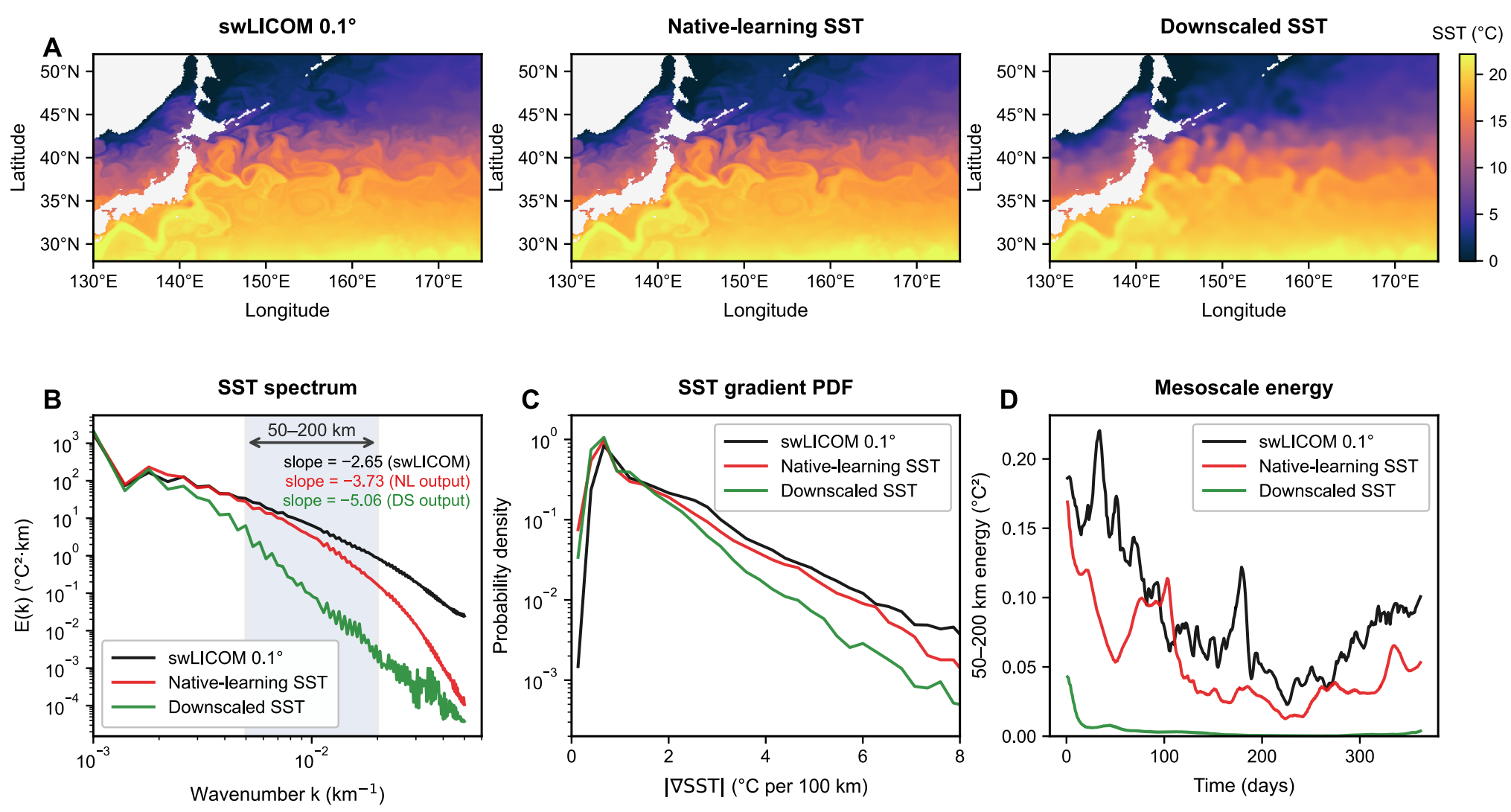

**Fig. 5. Comparison of native learning (NL) with downscaling (DS) in the Kuroshio Extension. (A)** SST snapshots from the swLICOM 0.1° simulation (left), Native-learning SST (middle), and Downscaled SST obtained by downscaling the coarse-grid 1° SST emulator output (right). **(B–D)** SST power spectra, spatial probability density functions of time-mean SST-gradient magnitude G, and the time evolution of 50–200 km mesoscale energy during the test period, respectively. In **(B)**, the gray band denotes wavelengths of 50–200 km, with fitted slopes of −2.65, −3.73, and −5.06 for swLICOM 0.1°, Native-learning SST, and Downscaled SST, respectively. Black, red, and green denote swLICOM 0.1°, Native-learning SST, and Downscaled SST throughout **(B–D)**. All three comparison fields are on the 0.1° grid. Although both AI pathways output fields on this grid, direct training retains substantially more high-wavenumber energy, persistently strong frontal gradients, and mesoscale variability. In **(C)**, temporal means cover the 2023 test period within 130–175° E, 28–52° N. PDFs use 45 bins over 0–12 °C per 100 km and are displayed over 0–8 °C per 100 km.

The quantitative diagnostics show that this smoothing corresponds to a substantial loss of mesoscale variance. The fitted spectral slope of Downscaled SST over 50–200 km was −5.06, markedly steeper than −2.65 for swLICOM 0.1° and −3.73 for the directly trained emulator (Fig. 5B). Its high-wavenumber tail decayed rapidly, showing that writing the output on a 0.1° grid did not restore the missing small-scale variance. Correspondingly, the spatial distribution of time-mean SST-gradient magnitude shifted toward weaker gradients, with persistently strong gradients less represented than in Native-learning SST (Fig. 5C). The 50–200 km energy of downscaled SST also decreased rapidly after initialization and remained close to zero during most of the test period (Fig. 5D).

Although the native 0.1° SST emulator still underestimated part of the mesoscale energy and retained a spectrum that was too steep, its output outperformed downscaled SST in Kuroshio high-wavenumber spectral power, the spatial occurrence of persistently strong gradients, and band-integrated energy. Because both AI pathways produced fields on the same 0.1° grid, the difference cannot be explained by output grid spacing alone. Within the tested configurations, the pathway that learned temporal evolution directly in the high-resolution state space retained greater mesoscale fidelity in the Kuroshio Extension. Across the additional regions, Native-learning SST generally retained more mesoscale variance, although the ranking varied by diagnostic (Fig. S5). In the weak-variability South Pacific subtropical gyre, downscaling yielded a closer spectral slope, but both pathways strongly underestimated band energy (Fig. S5, P–R).

### Mesoscale statistics are reproducible across diverse hardware platforms

Reproducibility across diverse hardware platforms is an important prerequisite for the scientific application of an ocean emulator. However, for a chaotic ocean system integrated autoregressively over long periods, reproducibility should not be equated with strict bitwise identity. Small differences in hardware architecture, numerical operator implementation, and floating-point rounding can amplify over time, causing trajectories generated from the same model weights to gradually diverge at the pointwise level *(22, 23)*. We therefore did not require the emulator outputs to remain identical at every grid point and time step. Instead, we assessed whether the statistical and physical characteristics relevant to the scientific conclusions of this study remained reproducible across hardware and software configurations.

The same native 0.1° SST emulator weights were evaluated on multiple computing platforms, including an NVIDIA H800 GPU, a Hygon DCU, a Hygon CPU, an NVIDIA RTX 5090, and an Intel x86 CPU. The SST root-mean-square error (RMSE) curves relative to swLICOM 0.1° were nearly indistinguishable across platforms, reproducing the same rapid initial growth and subsequent slower evolution (Fig. 6A). Differences in inference implementation therefore did not alter the overall time dependence of model error.

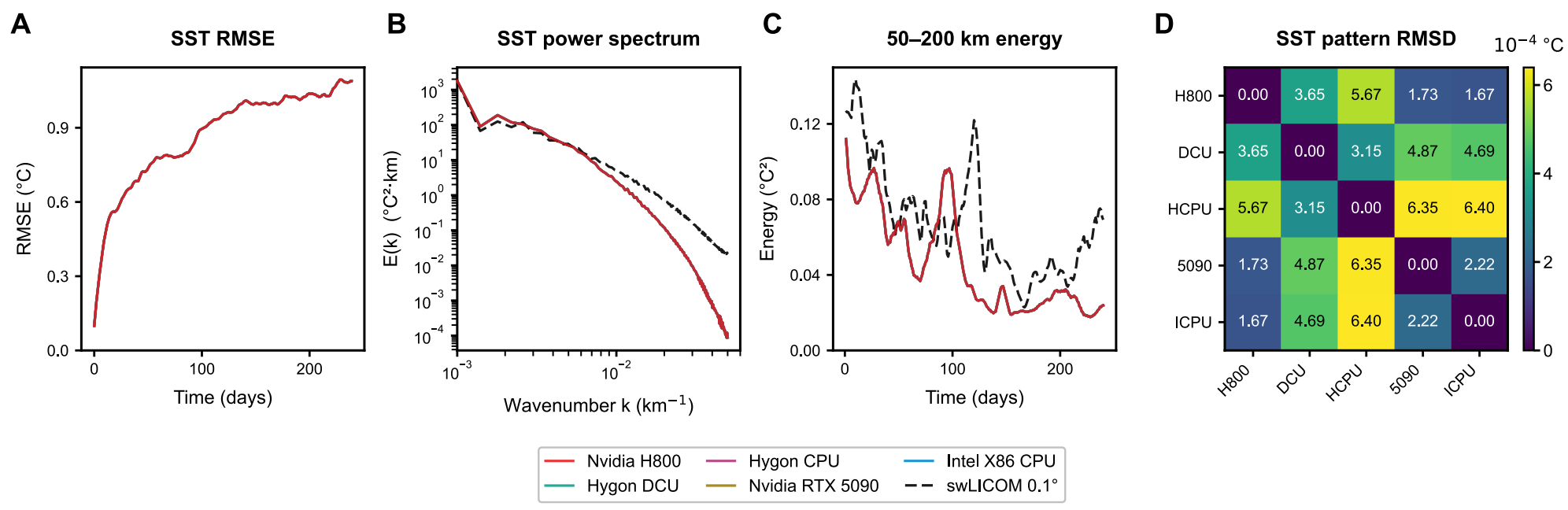

**Fig. 6. Statistical reproducibility of the fixed-weight native 0.1° SST emulator across diverse hardware platforms. (A)** SST RMSE through time for rollouts on an NVIDIA H800 reference platform (red), Hygon DCU (teal), Hygon CPU (purple), NVIDIA RTX 5090 (gold), and Intel x86 CPU (blue); all platform curves are solid. **(B)** Kuroshio Extension SST power spectra from the same hardware and software configurations, shown with the same colors; the black dashed line denotes swLICOM 0.1°. **(C)** Time evolution of 50–200 km mesoscale energy across platforms, again using the same colors; the black dashed curve denotes swLICOM 0.1°. **(D)** Pairwise SST pattern RMSD among platform outputs; cell annotations give the RMSD values and the color bar shows their magnitude.

The mesoscale diagnostics were similarly consistent. SST power spectra in the Kuroshio Extension overlapped across the full resolved wavenumber range, with no visible hardware-dependent displacement of the high-wavenumber roll-off (Fig. 6B). The 50–200 km energy curves followed the same sequence of peaks and troughs and retained similar overall energy levels, apart from small amplitude differences late in the rollout (Fig. 6C). Pairwise root-mean-square differences (RMSDs) between the mean SST patterns generated across platforms remained below $6.4 \times 10^{-4}$ °C, indicating nearly identical spatial patterns across hardware and software configurations (Fig. 6D).

## DISCUSSION

### Main findings and implications

We compared two pathways for obtaining eddy-resolving AI ocean simulations. Pathway 1 (NL) learned ocean-state evolution directly from 0.1° simulations, such that both temporal integration and the final output remained on the high-resolution grid. Pathway 2 (DS) first evolved SST with the coarse-grid 1° SST emulator and then mapped each output to the 0.1° grid using a separate deterministic downscaling module. The full-depth 1° emulator and native 0.1° SST emulator maintained stable large-scale climate baselines during long free-running simulations without substantial drift. These stability assessments provide complementary context for the evaluation of mesoscale statistics.

Although both pathways produced final outputs on the same 0.1° grid and were evaluated with identical mesoscale diagnostics, the emulator that learned dynamical evolution directly on the 0.1° grid better reproduced eddies, frontal meanders, and filaments and remained closer to the eddy-resolving reference in high-wavenumber spectra, mesoscale band energy, and spatial distributions of time-mean SST-gradient magnitude in the Kuroshio Extension. The pathway based on 1° evolution followed by downscaling recovered the large-scale position and morphology of the Kuroshio Extension, but produced smoother fields, a more rapidly decaying high-wavenumber spectrum, and lower spatial occurrence of persistently strong gradients and weaker 50–200 km variability. The time-mean diagnostics and energy evolution over most of the 2023 test period show that this contrast was not confined to individual snapshots.

These findings do not suggest that downscaling approaches are unnecessary, but rather highlight the fundamental distinction between generating high-resolution fields and learning high-resolution ocean dynamics. Our results demonstrate that, within the deterministic downscaling framework considered here, reconstructing a 0.1° field from coarse-resolution dynamical evolution is insufficient to fully recover the mesoscale statistics and dynamical characteristics. This underscores the continued importance of developing native high-resolution AI ocean models that directly learn the evolution of fine-scale dynamics. Nevertheless, future downscaling approaches may need to move beyond purely data-driven spatial reconstruction by incorporating physical constraints, dynamical consistency, or hybrid physics–AI formulations to more faithfully represent unresolved ocean processes.

### How mesoscale information is resolved by emulators

Our results raise the question of how mesoscale information is lost along the downscaling pathway. The rapid decay of the high-wavenumber spectral tail, the reduced spatial occurrence of persistently strong gradients, and the near disappearance of mesoscale band energy in Fig. 5 show that the difference is not merely a slight visual blurring of individual snapshots. It reflects a systematic change in the distribution of mesoscale variance. The present experiments, however, do not separate information loss during coarse-grid temporal evolution from limitations of the spatial mapping itself, and the observed discrepancy should not be assigned to a single mechanism without further tests. Even with this limitation, the results demonstrate that, within the deterministic pathway examined here, recovering high-resolution fields from coarse dynamics alone was insufficient to reproduce the statistical characteristics of native eddy-resolving simulations.

One possible explanation is that a 1° state does not uniquely constrain the phase and local organization of the corresponding 0.1° mesoscale field. The same coarse large-scale state may be compatible with different eddy positions, frontal roll-ups, and filamentary structures *(17, 24)*. Under a pointwise deterministic loss, the mapper may preferentially reconstruct large-scale components that are predictable from the coarse input while averaging high-frequency components that are not uniquely determined. This interpretation is consistent with the losses of spectral energy and gradient tails observed here, but it remains a hypothesis rather than a demonstrated mechanism.

A related question is whether stronger downscaling approaches can narrow the gap. Could temporal context, multiscale or spectral objectives, dynamical constraints, or joint training of the temporal and spatial modules retain more mesoscale variability *(12, 16, 17)*? What is the attainable performance of the mapper when it receives truly coarse-grained swLICOM states, and how much additional degradation occurs when it receives outputs from a free-running coarse emulator? Addressing these questions will require separating spatial-mapping error from coarse-rollout error and comparing multiple well-optimized representatives under matched data, computational budgets, and diagnostics. The present study is therefore best interpreted as identifying and defining a training-path problem, not as providing a final verdict on all possible downscaling methods.

**Scope of the deterministic comparison and generative downscaling**

The conclusion supported by this study applies to the specific experimental framework considered here: direct training at 0.1° achieved higher mesoscale fidelity than deterministic downscaling from coarse-resolution dynamics. This result should not be generalized to the claim that no downscaling method can recover mesoscale statistics. Rather, it highlights the resolution at which dynamics are learned as a key factor in high-resolution ocean emulation, while leaving open how future physics-aware downscaling methods may narrow this gap. A more definitive comparison would require competitive models for both pathways, comparable levels of hyperparameter optimization, and matched training data, computational budgets, diagnostics, and evaluation periods.

Generative downscaling offers a different way to formulate the unresolved mapping. Diffusion- or generative adversarial network (GAN)-based models can sample small-scale components that are not uniquely determined by a coarse input and may improve spectra, gradient distributions, or extreme-event frequencies at the ensemble level *(12, 17, 24)*. Whether this is sufficient for the problem considered here remains open. If an ensemble reproduces the reference marginal statistics, are the small-scale phases of individual members dynamically consistent with the large-scale state and neighboring time steps? Can stochastic structures evolve continuously rather than change discontinuously under framewise generation? When a generative mapper is coupled to a free-running coarse emulator, does mesoscale energy remain stable over long integrations?

These questions require several levels of evaluation to be distinguished. Spectral and probability-density agreement characterizes ensemble or marginal statistics; single-field agreement measures spatial correspondence for a particular state; temporal coherence tests consistency between consecutive states; and stable free-running integration addresses error accumulation and maintenance of the climate state. A visually sharp snapshot may satisfy only part of this hierarchy. Future evaluations of both deterministic and generative downscaling should therefore include power spectra, band-limited energy, gradient distributions, temporal consistency, and long free-running diagnostics *(8, 17, 24)*.

**Cross-platform statistical reproducibility and emulator-as-storage**

A practical motivation for developing AI ocean emulators is to enable the "emulator-as-storage" paradigm: instead of archiving petabyte-scale outputs from eddy-resolving simulations, trained emulator weights could serve as a compact and executable representation of the simulation, enabling efficient storage, dissemination, and on-demand regeneration of high-resolution ocean fields. Realizing this paradigm requires the emulator to produce statistically consistent results across diverse computing platforms, motivating the assessment of cross-platform reproducibility. Here, we evaluate reproducibility using multiple diagnostics, including error growth, spectral characteristics, mesoscale energy, and spatial patterns. Our results support these quantities as practical reproducibility targets for a fixed-weight emulator.

This statistical reproducibility provides an initial basis for emulator-as-storage. Instead of archiving every field from a high-resolution simulation, one could archive the model weights and inference configuration and regenerate statistically equivalent fields when needed *(11, 12)*. For the daily 0.1°

SST fields considered here, one 1800 × 3600 float32 field occupies approximately 25.9 MB, corresponding to about 950 GB for 100 years. In contrast, the trained emulator weights require 486 MB, which corresponds to a checkpoint-only storage reduction of 99.95% per century. More importantly, this approach could enable wider dissemination of eddy-resolving simulations: users with suitable local computing resources could deploy the emulator rather than relying on large-scale supercomputing facilities.

In practice, the storage advantage does not by itself establish an end-to-end efficiency advantage. Regeneration also requires initial fields, preprocessing metadata, the software environment, and any forcing data, while total wall-clock time depends on inference and output-writing costs. Exact archived trajectories may remain necessary for reproducibility or analyses that cannot use statistically equivalent regeneration. Emulator-as-storage should therefore be treated as a testable data-distribution concept rather than a demonstrated replacement for conventional archives *(25, 26)*.

**Limitations and future directions**

The controlled comparison and mesoscale diagnostics in this study focus on SST. SST clearly exposes fronts and part of the mesoscale spatial variability, but it does not fully describe the ocean dynamical state *(21, 27, 28)*. The same comparison should be extended to sea surface height, horizontal currents, salinity, and three-dimensional temperature, salinity, and velocity fields. It will then be possible to determine whether improved surface spatial structures and spectral statistics correspond to credible pressure fields, current structures, and subsurface responses. Eddy kinetic energy, heat and tracer transport, frontal position, eddy lifetime and propagation, and extreme-gradient events would provide complementary tests of whether visually identifiable eddies exert the correct dynamical effects *(1, 3, 4)*.

In contrast to the 1° emulator, which closely tracks the reference simulation, the native 0.1° emulator exhibits a systematic cold bias with coherent large-scale patterns, particularly across the tropical Pacific (Figures 2 and S1). This bias may partly arise from differences in temporal sampling. The 1° emulator learns from monthly data that filter high-frequency variability, whereas the 0.1° emulator learns from daily data and must capture both short-term variability and slower background evolution. The reduced RMSE after 5-day averaging supports this interpretation, suggesting that high-frequency variability contributes to the performance gap (Figure S6). Future work will explore multiscale learning to better represent variability across temporal scales and improve long-term stability.

The emulators in this study are trained and evaluated against dynamics encoded by swLICOM. The conclusions therefore address fidelity to a particular numerical model and do not automatically imply accuracy relative to the real ocean *(8, 30, 31)*. Applying the same controlled training-path comparison to other eddy-resolving numerical models, and validating independently against satellite SST, sea surface height, and drifting-buoy observations, would provide a direct test of transferability *(21, 27, 28)*. Would different reference models yield the same advantage for direct learning? Would observational constraints alter the assessment of spectral slopes, frontal intensity, or mesoscale energy? These questions can be tested without changing the basic experimental framework developed here.

Last, the present comparison is based on one set of deterministic trainings and matched initial conditions. Whether different random seeds, initialization dates, and small initial perturbations alter the statistical separation between the pathways remains to be assessed with multiple realizations and ensemble experiments. Future work should distinguish model uncertainty, internal variability, and training uncertainty, and quantify the pathway-to-pathway differences relative to these uncertainty ranges *(12, 24, 29)*.

Overall, this study shows that the credibility of a high-resolution AI ocean field cannot be inferred from low-resolution dynamics followed by downscaling reconstruction alone *(6, 13, 17)*. In the present controlled experiments, learning temporal evolution directly in an eddy-resolving state space preserved mesoscale statistics better than coarse evolution followed by deterministic spatial reconstruction. The broader questions remain open: Which coarse-state information is sufficient to constrain dynamically consistent small-scale structures? Which model designs can recover that information? And which diagnostics are needed to establish that the recovered fields are suitable for scientific analysis and climate assessment? Answering these questions will be central to moving high-resolution AI ocean simulation from generating sharp fields to generating trustworthy scientific data.

## MATERIALS AND METHODS

### 1. Experimental design

We designed a controlled comparison to test whether mesoscale sea-surface temperature (SST) evolution is represented more faithfully when temporal dynamics are learned directly on the native 0.1° grid or first learned on a 1° grid and then reconstructed at 0.1° by deterministic downscaling (Fig. 1). Both pathways were derived from the same daily 0.1° swLICOM SST dataset. In the native learning pathway (NL), the native 0.1° SST emulator predicted successive 0.1° SST states. In the downscaling pathway (DS), the same fields were conservatively coarse-grained to 1°, evolved by the separate coarse-grid 1° SST emulator, and mapped frame by frame back to 0.1° by a downscaling module. Here, "native" denotes learning in the high-resolution state space on the remapped regular grid described in Section 2.2, rather than on the original tripolar grid.

The two pathways used the same 2013–2022 source period and were initialized from temporally matched states for evaluation from January to December 2023. Their outputs were compared with the same native 0.1° swLICOM fields on the same dates and grid using identical diagnostics. The comparison therefore measures the combined consequence of learning temporal evolution on a coarse grid and subsequently reconstructing spatial detail, relative to learning the evolution directly on the native grid. Separate rollouts of the full-depth 1° emulator and the native 0.1° SST emulator were used to evaluate resolution-specific climate stability and are described in Sections 3 and 5. The full-depth 1° emulator is distinct from the daily coarse-grid 1° SST emulator and is not part of the DS processing chain.

### 2. Datasets

### 2.1 Full-depth 1° dataset

The full-depth 1° dataset comprised 124 model years of monthly mean output from a global LICOM3.0 integration conducted under the Ocean Model Intercomparison Project phase 1 (OMIP1) protocol *(18, 32, 33)*. Following Yu et al., CORE-II atmospheric forcing for 1948–2009 was repeated for two 62-year cycles, and the integration was initialized from PHC3 temperature and salinity.

The nominal horizontal resolution was 1°, and the vertical coordinate contained 30 levels extending from 5 to 5243 m. The prognostic state included potential temperature, salinity, zonal velocity, and meridional velocity at all 30 depths, together with sea surface height. Four prescribed surface-boundary fields—zonal surface ocean stress, meridional surface ocean stress, net heat flux, and net heat-flux anomaly—were supplied at every autoregressive update.

The original model grid was remapped with periodic longitude to a regular 180 × 360 latitude–longitude grid using bilinear remapping.

### 2.2 Eddy-resolving 0.1° SST dataset

The eddy-resolving 0.1° SST dataset was obtained from the output of a global high-resolution LICOM3.0 integration *(19, 34)* forced by JRA55-do atmospheric reanalysis *(35, 36)*. The native tripolar Arakawa-B grid contained 3600 × 2302 horizontal points and 55 vertical levels; the two northern grid poles were located near (65° N, 65° E) and (65° N, 115° W).

Daily output from 2013–2023 was used, and this study retained temperature at the uppermost 1.5-m level as SST. The native surface field was remapped with Climate Data Operators *(37)* to a regular 0.1° grid containing 1800 × 3600 points using bilinear remapping.

### 2.3 Coarse-grid 1° SST dataset

To construct paired data for the downscaling pathway, each remapped 0.1° SST field was conservatively remapped to a 1° regular grid of 180 × 360 points. The operation preserved the daily sampling and timestamps. The coarse and native datasets therefore differed only in spatial resolution and their associated normalization statistics. The resulting 1° fields were used both to train the single-variable coarse-grid 1° SST emulator and as the low-resolution inputs paired with native 0.1° targets during downscaling training.

## 3. Autoregressive ocean emulators

### 3.1 Network architecture

All autoregressive emulators used a convolutional U-Net based on the Samudra architecture *(8, 38)*. Four encoder stages combined modified ConvNeXt-style residual blocks with 2 × 2 average pooling *(39)*. A bottleneck block operated at the coarsest scale, and the decoder restored spatial resolution by bilinear upsampling and additive skip connections. Each ConvNeXt-style block used two 3 × 3 convolutions, a fourfold inverted channel expansion, batch normalization, capped Gaussian error linear unit activations (GELU clipped at 10) *(40)*, and a 1 × 1 projection within a residual connection. Dilations across the four encoder stages were 1, 2, 4, and 8. Longitude was padded periodically, latitude was padded with zeros, and a final 3 × 3 convolution produced the two future states.

The full-depth 1° emulator predicted 121 variables per time level: four three-dimensional variables at 30 depths and one sea-surface-height field. Its input therefore contained 242 prognostic channels from two consecutive months plus four boundary channels (246 channels total), and its output contained 242 channels for the following two months. Encoder widths were 200, 250, 300, and 400 channels.

The native 0.1° SST emulator predicted SST only. Two consecutive daily SST fields formed its two-channel input and the following two fields formed its two-channel output. Encoder widths were 50, 100, 200, and 250 channels. The coarse-grid 1° SST emulator used the same temporal formulation and channel widths but operated on the conservatively coarse-grained SST fields. Neither SST-only emulator received an external forcing channel.

Following the Samudra autoregressive framework *(8)*, each emulator used two consecutive standardized states to predict the next two states (Eq. 1):

$$(\hat{x}_{t+2\Delta t}, \hat{x}_{t+3\Delta t}) = F_\theta(x_t, x_{t+\Delta t}, b_{t+\Delta t}) \tag{1}$$

Here, *x* is the standardized ocean state, a hat denotes a prediction, *t* denotes time, $F_\theta$ is the emulator with learned parameters *θ*, and *b* contains the prescribed boundary fields. Training used overlapping 10-state windows: two swLICOM history states followed by four two-state autoregressive passes, with gradients propagated through all passes. The full-depth emulator received the prescribed boundary variables at each pass, whereas the SST-only emulators used no boundary input. The interval $\Delta t$ was one month for the full-depth 1° emulator and one day for both SST emulators. During inference, each emulator was initialized with two consecutive swLICOM states and rolled out freely.

The loss was the mean of per-channel mean-squared errors over the four autoregressive passes (Eq. 2):

$$\text{loss} = \frac{1}{4C}\sum_{s=1}^{4}\sum_{c=1}^{C}\left[\frac{1}{BHW}\sum_{i,j,k}\left(\hat{x}_{i,c,j,k}^{(s)} - x_{i,c,j,k}^{(s)}\right)^2\right] \tag{2}$$

Here, *s* indexes the four autoregressive passes; *c* indexes the *C* output channels, including both predicted time levels; and *i*, *j*, and *k* index the *B* batch samples, *H* latitude points, and *W* longitude points, respectively. The hatted and unhatted *x* values denote standardized predictions and swLICOM targets at pass *s*.

### 3.2 Training details

All three autoregressive emulators were optimized with Adam *(41)* using an initial learning rate of $2 \times 10^{-4}$. Per-process batch sizes were 4 for the full-depth 1° and coarse-grid 1° SST emulators and 1 for the native 0.1° SST emulator.

All three autoregressive emulators were trained on four NVIDIA H800 GPUs with 80 GB of memory per GPU.

## 4. Downscaling pathway

In the downscaling pathway, the coarse-grid 1° SST emulator first generated the daily SST sequence on the 1° grid. Each predicted 1° field was then passed independently to the deterministic downscaling module to obtain the corresponding 0.1° field. This formulation follows the general use of convolutional downscaling models to learn spatial mappings between coarse- and fine-resolution geophysical fields *(42, 43)*. The downscaling module had no recurrent state and therefore did not determine the temporal evolution; it only mapped each coarse emulator output to the native-resolution grid. The spatial mapping was defined as (Eq. 3):

$$\hat{y} = \mathrm{Bilinear}_{1^\circ \rightarrow 0.1^\circ}(x) + R_{\phi}(x) \tag{3}$$

Here, $x$ is the coarse SST input, $\hat{y}$ is the downscaled prediction, Bilinear denotes bilinear interpolation, and $R_{\phi}$ is the residual network with learned parameters $\phi$. The network learned a residual correction to a conventionally interpolated coarse field, following the residual-learning strategy commonly used in image super-resolution *(44, 45)*. The bilinear branch directly interpolated the coarse SST field from 1° to 0.1°. In parallel, the learned residual branch projected the single coarse input channel to 128 features with a 3 × 3 convolution and processed them with four ConvNeXt-style blocks on the 180 × 360 grid. A 1 × 1 convolution reduced the representation to 16 channels before bilinear interpolation to 1800 × 3600. Two additional ConvNeXt-style blocks refined the high-resolution features, and a final 3 × 3 convolution produced a one-channel residual that was added to the bilinear field. The blocks used batch normalization, capped GELU, non-dilated 3 × 3 convolutions, fourfold channel expansion, periodic longitude padding, and zero latitude padding.

The downscaling module was trained on time-matched pairs of standardized coarse 1° SST and native 0.1° SST fields. Data from 2013–2022 were used for training. Training followed the optimization and hardware configuration described in Section 3.2, with a per-process batch size of 4. The objective was mean-squared error in the standardized native-grid target space.

## 5. Rollout experiments

All rollouts were initialized with two consecutive swLICOM states. Thereafter, prognostic inputs were supplied exclusively by the preceding emulator outputs. The experiments are summarized in Table 1.

**Table 1. Rollout experiments and their roles in the analysis.**

| System | Experiment | Evaluation interval | Purpose |
|---|---|---|---|
| Full-depth 1° emulator | Test rollout | Years 95–124, monthly | 1° Climate baseline |
| | 100-year rollout | Years 95–194, monthly | 1° Long-term stability |
| Native 0.1° SST emulator | Test rollout | Year 2023, daily | 0.1° Climate baseline / NL evaluation |
| | 100-year rollout | Years 2023–2122, daily | 0.1° Long-term stability |
| Downscaled SST | Test rollout | Year 2023, daily | DS pathway evaluation |

For the full-depth 1° emulator, years 95–124 constituted the first 30 years of the same 100-year rollout spanning years 95–194; this test window was not a separately initialized experiment. The controlled comparison between direct high-resolution learning and coarse-resolution evolution with downscaling reconstruction used 2023 as an independent test year. For the centennial full-depth 1° rollout, the prescribed boundary forcing was repeated on a 30-year cycle.

## 6. Evaluation of climate stability and mesoscale structure

### 6.1 Global-mean SST and long-term trends

Global-mean SST was computed as the area-weighted spatial average over all ocean grid cells. A linear trend was obtained by ordinary least squares against elapsed time and reported in degrees Celsius per century.

### 6.2 Analysis regions

Mesoscale diagnostics were evaluated in six dynamically distinct regions using separate interior boxes for spectral calculations and larger display boxes for maps and gradient distributions: Kuroshio

Extension, 145–170° E, 30–42° N and 130–175° E, 28–52° N; Gulf Stream, 295–320° E, 35–45° N and 280–325° E, 30–50° N; Agulhas, 15–40° E, 35–45° S and 5–45° E, 30–50° S; Brazil–Malvinas, 300–315° E, 38–48° S and 290–325° E, 33–53° S; Antarctic Circumpolar Current, 200–230° E, 48–58° S and 190–240° E, 44–62° S; and the South Pacific subtropical gyre, 210–240° E, 20–30° S and 200–250° E, 10–40° S.

### 6.3 Radial SST spectra and band-limited mesoscale variance

Here, $F$ denotes the Fourier coefficients, $|F|^2$ their squared magnitude, $k$ the radial wavenumber in $km^{-1}$, and $E(k)$ the corresponding one-dimensional spectral density. For each time and region, the spatial mean was removed from the two-dimensional SST field and a separable two-dimensional Hann window was applied. We computed the two-dimensional discrete Fourier transform, normalized $|F|^2$ by the number of grid points and the squared-window energy, and radially binned the spectrum in wavenumber increments set by the longest side of the regional array. Wavenumbers above the single-axis Nyquist frequency were discarded. The azimuthal spectral density was multiplied by the analytical annulus circumference to produce a one-dimensional spectrum $E(k)$ in °C² km, for which $\int E(k)\ dk$ approximates spatial SST variance.

The spectral slope was estimated by ordinary least squares of $\log_{10} E(k)$ on $\log_{10} k$ over wavelengths of 50–200 km, using the time-mean spectrum and requiring at least three positive finite bins. The 50–200 km band-limited SST variance at time $t$ was defined as (Eq. 4)

$$E_{50-200}(t) = \int_{50}^{200} E_\lambda(\lambda, t)\ d\lambda \tag{4}$$

where λ is wavelength in kilometres and $E_\lambda$ is the corresponding wavelength-domain spectral density. At least two spectral bins were required for numerical integration. The 1° fields did not meet this criterion in the 50–200 km band and were therefore reported as unresolved rather than assigned zero variance. Where centennial 0.1° spectra were analysed, one field was sampled every five days before calculating the regional spectra.

### 6.4 SST gradients and variability

SST-gradient magnitude was calculated on the sphere at every time step. With Earth radius $R$ = 6371 km, latitude $\varphi$, and longitude $\lambda$ (Eq. 5),

$$|\nabla T| = \sqrt{\left(\frac{1}{R\cos\phi}\frac{\partial T}{\partial \lambda}\right)^2 + \left(\frac{1}{R}\frac{\partial T}{\partial \phi}\right)^2} \tag{5}$$

Here, $T$ is SST, ∇ is the horizontal gradient operator, and ∂ denotes a partial derivative; angular coordinates in Eq. 5 are in radians. In $G(x)$, $x$ denotes spatial position and $\text{mean}_t$ denotes the temporal mean. Derivatives were evaluated with centered finite differences using the actual coordinate spacing and converted to degrees Celsius per 100 km. Gradients crossing land were excluded through propagation of the swLICOM wet mask. The principal gradient map was the temporal mean of the instantaneous gradient magnitude, $G(x) = \text{mean}_t(|\nabla \text{SST}(x, t)|)$. Means used the available daily fields in the 2023 test rollout at 0.1° and monthly fields from model years 95–124 at 1° (Table 1). Regional probability densities were formed from the finite grid-point values of this time-mean map within each display box, using 45 equally spaced bins from 0 to 12 °C per 100 km for the 0.1° fields. The archived 1° distributions used 30 bins from 0 to 8 °C per 100 km, with the same bin width. All gradient PDFs are displayed from 0 to 8 °C per 100 km; this plotting limit does not identify the maximum gradient. These distributions therefore quantify the spatial occurrence of persistently strong frontal gradients and are not probability densities of all instantaneous gradient events.

For supplementary variability maps (Fig. S4, A–C), SST anomalies were obtained by subtracting a centered 30-sample running mean at every grid point, after which the temporal standard deviation was calculated. The corresponding frontal-gradient maps show $G$ for the same test period (Fig. S4, D–F).

### 6.5 RMSE, temporal R², and pairwise RMSD

We used the cosine-latitude-weighted spatial root-mean-square error (RMSE) to quantify the difference between the simulated SST and the reference field (Eq. 6):

$$\mathrm{RMSE}(t) = \sqrt{\frac{\sum_i w_i[\hat{x}_i(t) - x_i(t)]^2}{\sum_i w_i}}, \quad w_i = \cos\phi_i \tag{6}$$

Here, $i$ indexes ocean grid cells; $\hat{x}_i(t)$ and $x_i(t)$ are simulated and reference SST at time $t$, respectively; and $w_i$ is the cosine-latitude weight at latitude $\phi_i$.

At each ocean grid point, the temporal coefficient of determination was calculated as $R^2 = 1 - \frac{\sum_t(\hat{x}_i(t)-x_i(t))^2}{\sum_t(x_i(t)-\bar{x}_i)^2}$, where $\hat{x}_i(t)$ and $x_i(t)$ denote simulated and reference SST at time $t$, respectively, and $\bar{x}_i$ is the reference mean at grid point $i$ over paired valid times. Calculations used SST without removing the seasonal cycle; grid points with fewer than two valid pairs or constant reference SST were excluded.

The root-mean-square difference (RMSD) between each pair of computational platforms was calculated using Eq. 6, with ($\hat{x}_i(t)$) and ($x_i(t)$) replaced by the time-mean SST fields produced by the two platforms over the common 240-frame comparison window (Section 7.1).

## 7. Cross-platform inference and statistical reproducibility

### 7.1 Controlled inference experiments

Numerical reproducibility was assessed by applying the same final 0.1° SST emulator checkpoint, evaluation interval, and two initial SST fields on five computing platforms: an NVIDIA H800 reference platform, a Hygon DCU, a Hygon CPU, an NVIDIA RTX 5090, and an Intel x86 CPU. The emulator received no external forcing. Because the platforms differed substantially in throughput and available resources, the first 240 daily output fields were used as the common comparison window. This choice ensured identical dates and sample size across platforms and was made for computational comparability rather than to represent a dynamical timescale.

For each platform, the global SST RMSE relative to swLICOM, Kuroshio radial spectra, and 50–200 km band-limited variance time series were calculated with the procedures above. Pairwise RMSDs were calculated from the 240-frame mean SST maps. Bitwise identity was not required because small platform-dependent floating-point differences can lead to trajectory divergence in an autoregressive chaotic system; the experiment instead tested reproducibility of error growth, spatial means, spectra, and band-limited variance.

## Declarations

**Competing interests:** The authors declare that they have no conflict of interest.

**Data, code, and materials availability:** Data and code supporting this study will be made publicly available upon journal publication.

## Supplementary Materials

Figs. S1 to S6

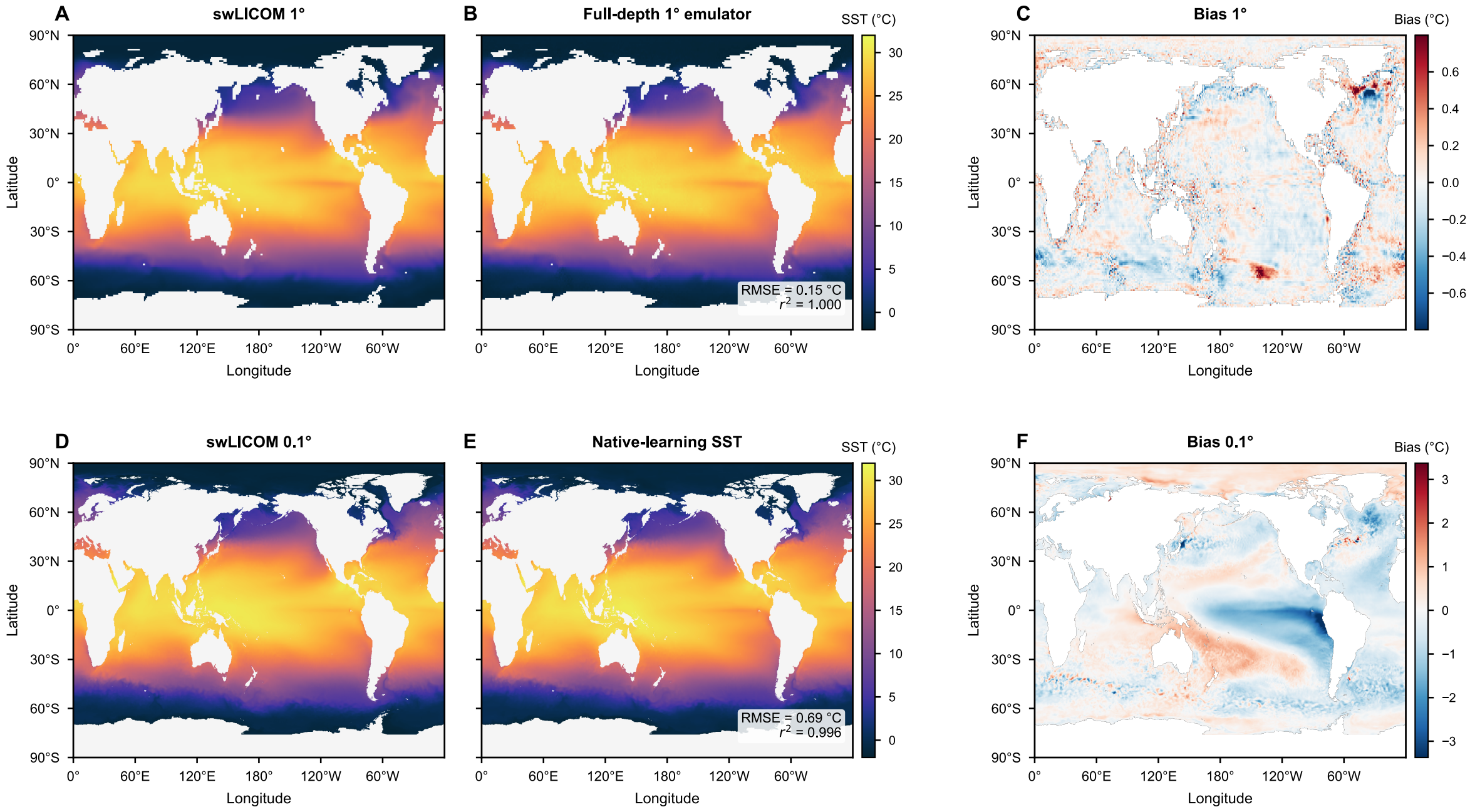

**Fig. S1. Global SST climatology and bias for the full-depth 1° emulator and native 0.1° SST emulator. (A–C)** swLICOM 1° climatology, Full-depth 1° emulator climatology, and the full-depth 1° emulator-minus-swLICOM bias, respectively. **(D–F)** swLICOM 0.1° climatology, Native-learning SST climatology, and the corresponding 0.1° bias. SST and bias use the color scales shown in the figure, and white areas denote land. In (B, E), RMSE and $r^2$ are calculated from the time-mean SST maps using spatial weighting; $r^2$ is the square of the weighted spatial Pearson correlation coefficient.

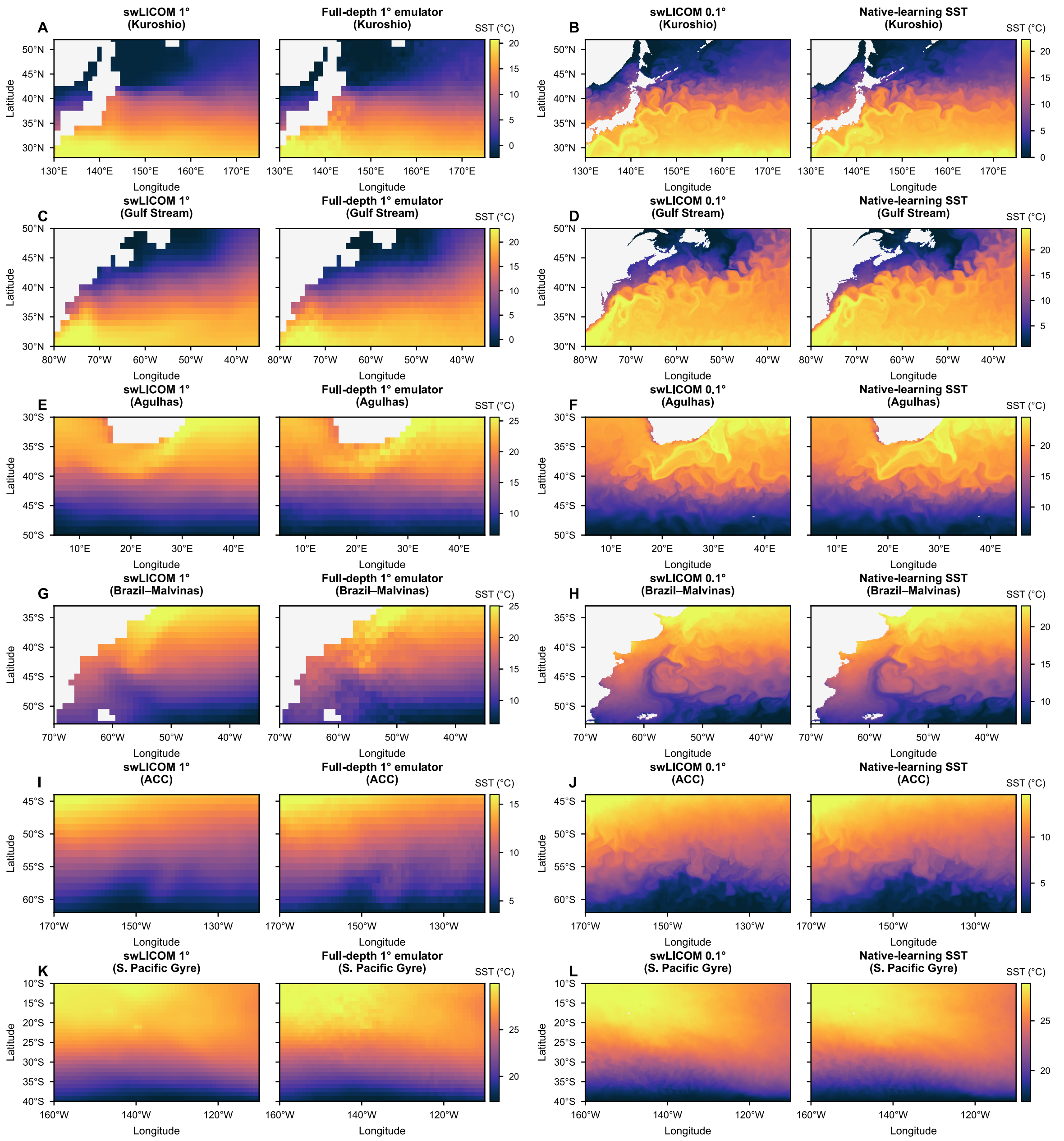

**Fig. S2. Regional SST snapshots at 1° and 0.1° in multiple dynamically active regions.** Within each lettered panel, swLICOM is shown on the left and the corresponding emulator on the right. **(A, B)** Kuroshio Extension; **(C, D)** Gulf Stream; **(E, F)** Agulhas region; **(G, H)** Brazil–Malvinas Confluence; **(I, J)** Antarctic Circumpolar Current; and **(K, L)** South Pacific subtropical gyre. Odd-lettered panels show 1° fields from swLICOM and the full-depth 1° emulator; even-lettered panels show 0.1° fields from swLICOM and the native 0.1° SST emulator (Native-learning SST). The comparisons assess regional consistency of broad frontal structures and, at 0.1°, eddies, frontal meanders, and filaments.

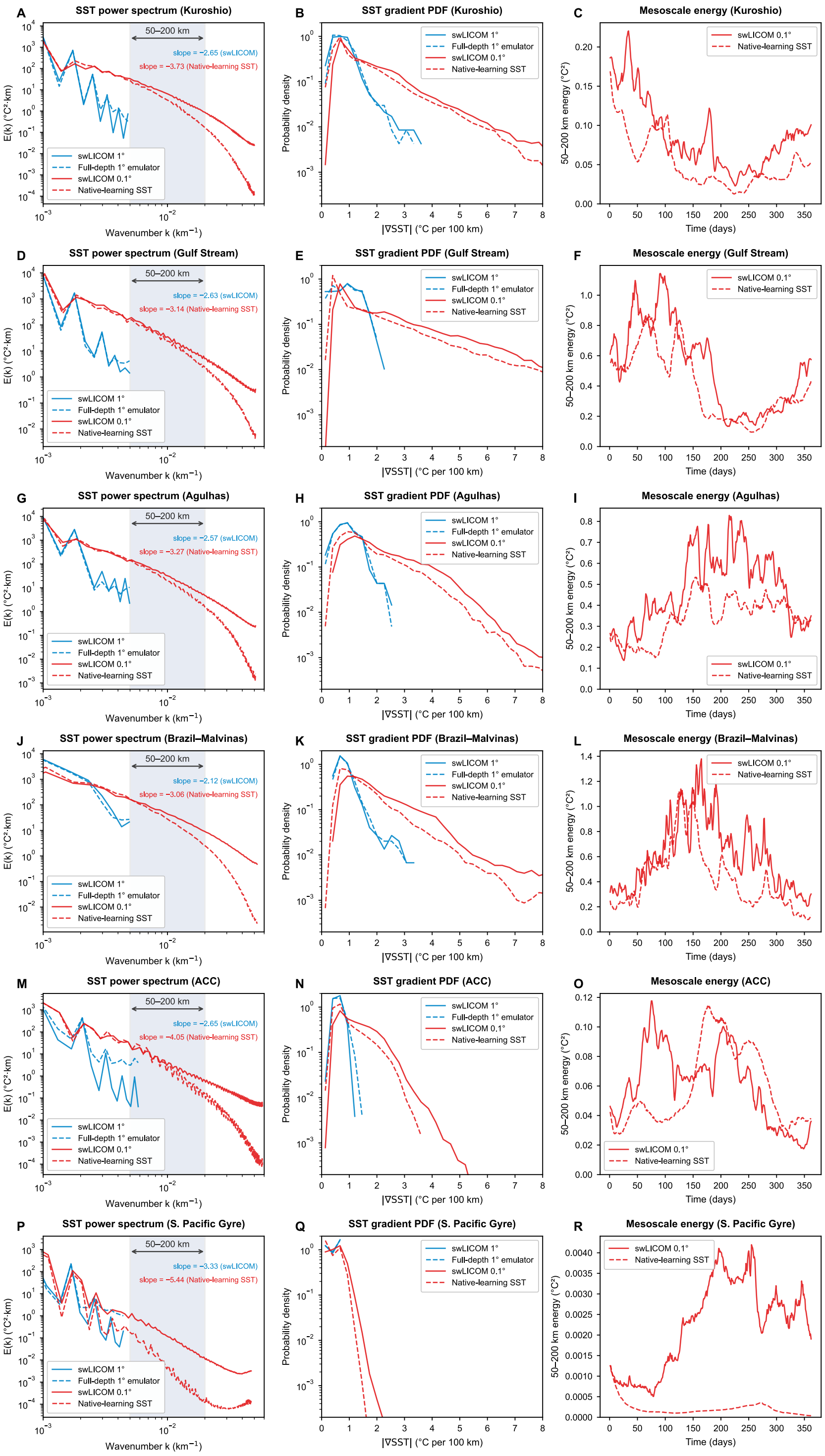

**Fig. S3. Regional mesoscale diagnostics for the full-depth 1° emulator and native 0.1° SST emulator.** Rows show **(A–C)** Kuroshio, **(D–F)** Gulf Stream, **(G–I)** Agulhas, **(J–L)** Brazil–Malvinas, **(M–O)** Antarctic Circumpolar Current, and **(P–R)** South Pacific subtropical gyre. The left column shows SST power spectra, with the 50–200 km band shaded and fitted slopes annotated; the middle column shows spatial probability density functions of time-mean SST-gradient magnitude G; and the right column shows the time evolution of 50–200 km mesoscale energy for swLICOM 0.1° and Native-learning SST. Blue denotes 1° and red denotes 0.1°, with solid and dashed line styles identified in each panel legend. Gradient means cover model years 95–124 at 1° and 2023 at 0.1°, using each regional display box in Methods Section 6.2. The 1° and 0.1° curves use 30 bins over 0–8 and 45 bins over 0–12 °C per 100 km, respectively; both are displayed over 0–8 °C per 100 km.

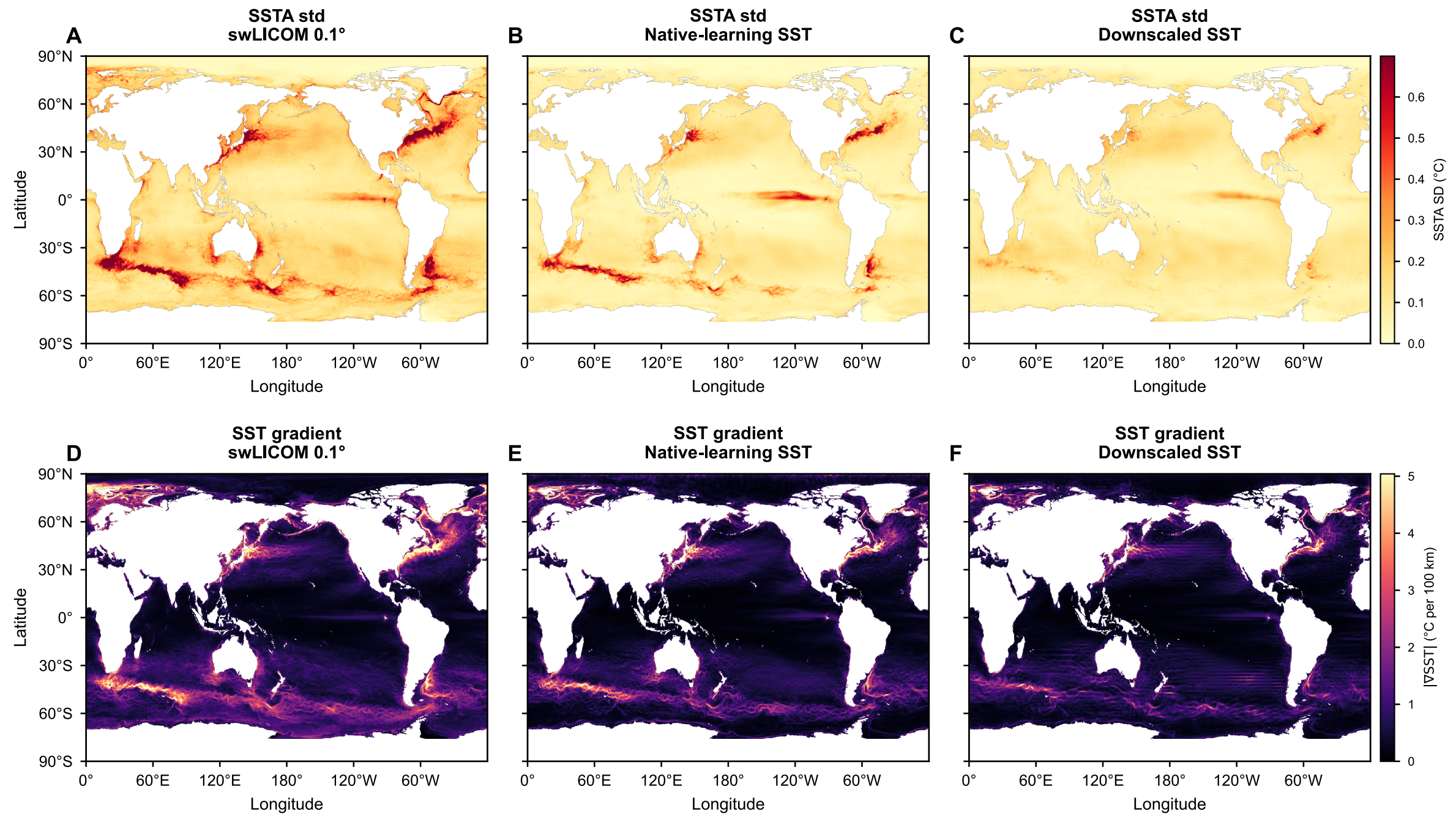

**Fig. S4. Global SST variability and frontal gradients for native learning (NL) and downscaling (DS). (A–C)** Standard deviation of SST anomalies for swLICOM 0.1°, Native-learning SST, and Downscaled SST, respectively. **(D–F)** Time-mean SST-gradient magnitude, G, for the same three datasets over the 2023 test period. The native 0.1° SST emulator better retains the concentration of high variability and strong gradients along western boundary currents and the Antarctic Circumpolar Current, whereas the downscaled field is smoother overall. White areas denote land. All three comparison fields are on the 0.1° grid.

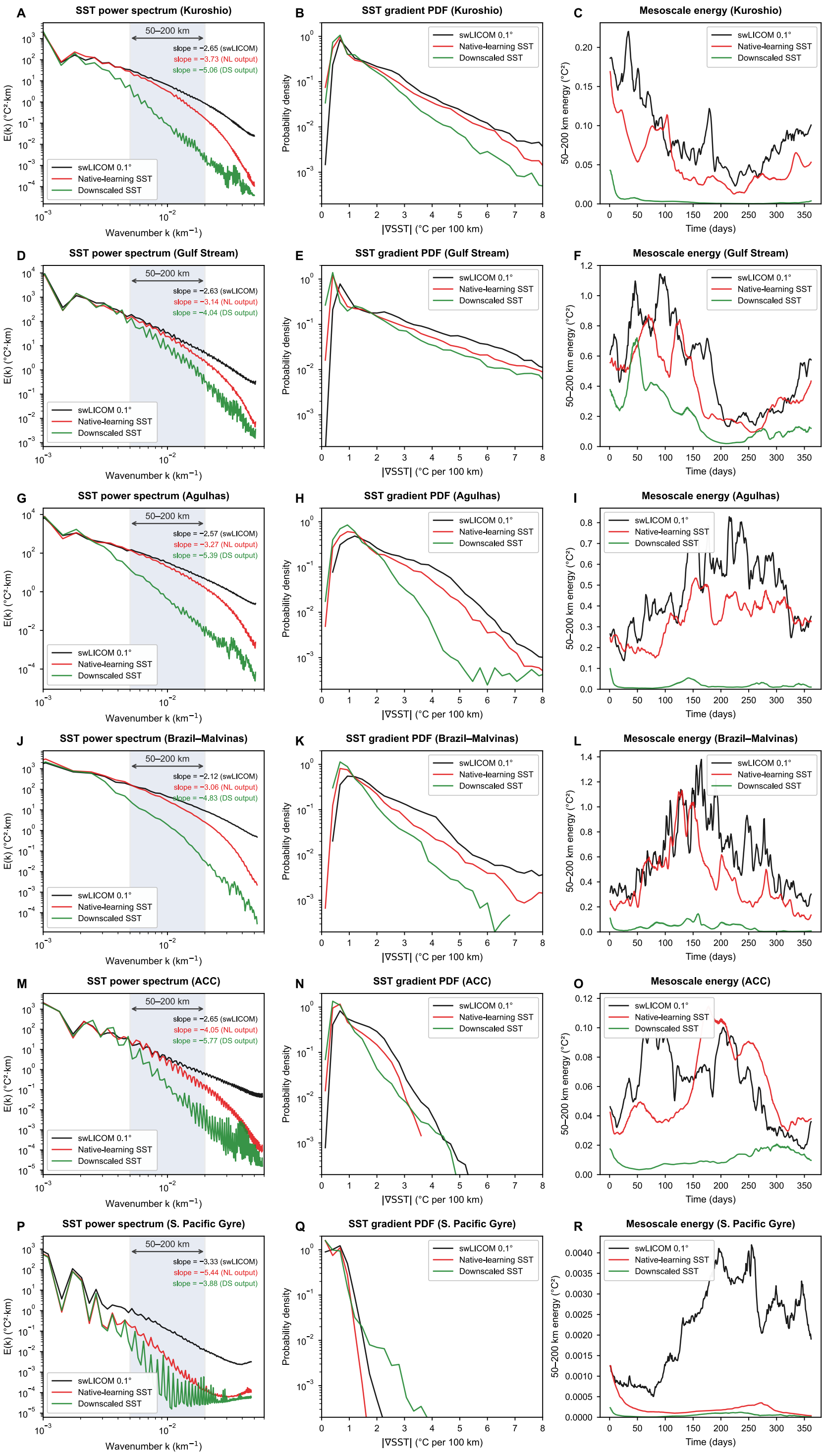

**Fig. S5. Regional mesoscale comparison of native learning (NL) with downscaling (DS).** Rows show **(A–C)** Kuroshio, **(D–F)** Gulf Stream, **(G–I)** Agulhas, **(J–L)** Brazil–Malvinas, **(M–O)** Antarctic Circumpolar Current, and **(P–R)** South Pacific subtropical gyre. The left column shows SST power spectra, with the 50–200 km band shaded and fitted slopes annotated; the middle column shows spatial probability density functions of time-mean SST-gradient magnitude G; and the right column shows the time evolution of 50–200 km mesoscale energy. Black, red, and green denote swLICOM 0.1°, Native-learning SST, and Downscaled SST, respectively. All three comparison fields are on the 0.1° grid. Gradient means cover the 2023 test period within each regional display box in Methods Section 6.2. PDFs use 45 bins over 0–12 °C per 100 km and are displayed over 0–8 °C per 100 km.

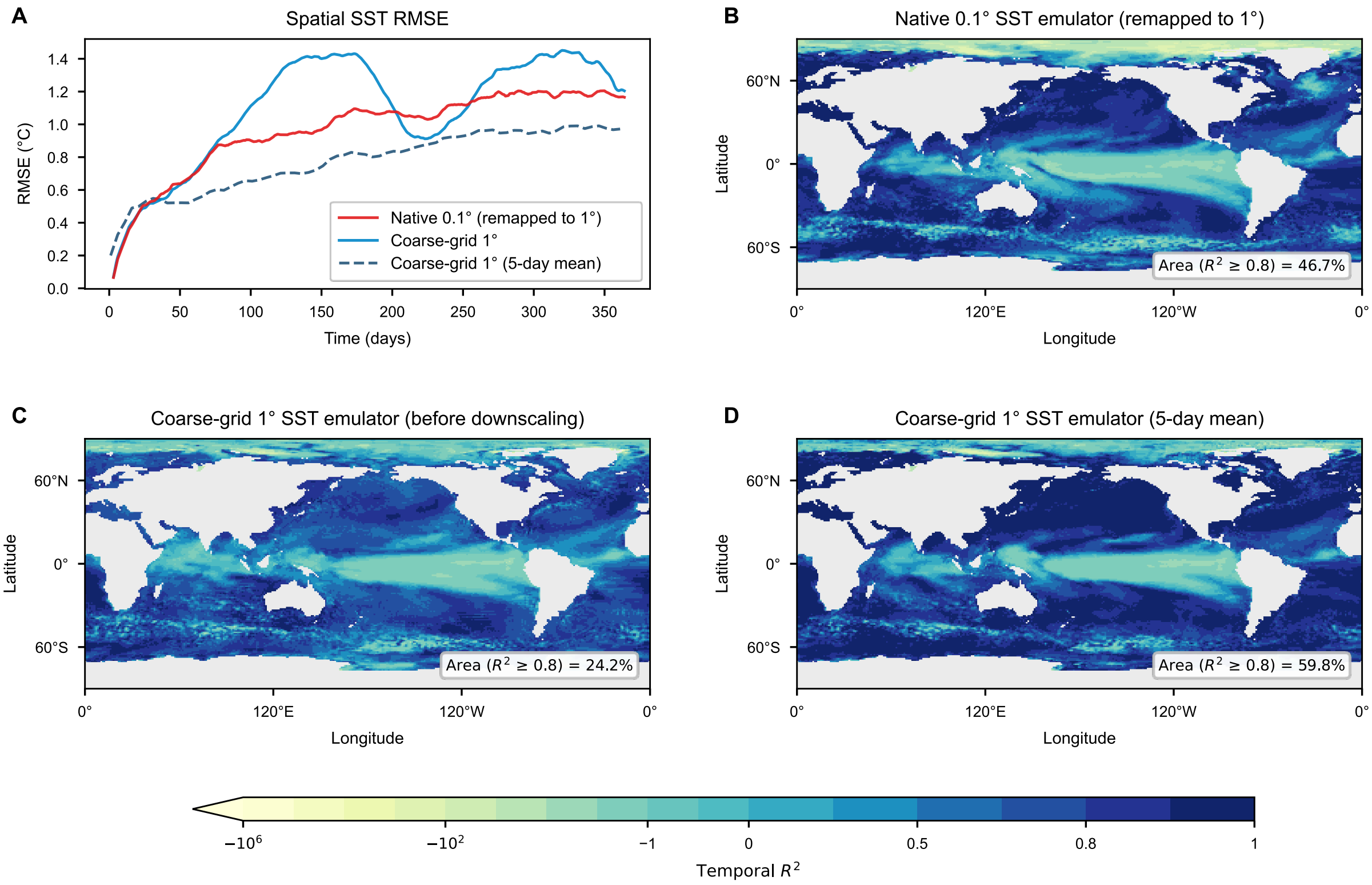

**Fig. S6. Pointwise SST agreement on a common 1° grid.** (A) Spatial SST RMSE for the native 0.1° SST emulator conservatively remapped to 1°, the coarse-grid 1° SST emulator, and the five-day-mean 1° SST emulator during the 2023 evaluation period. (B–D) Grid-point temporal $R^2$ for the same configurations, respectively. The five-day-mean configuration is trained and evaluated using five-day-mean SST fields, with predictions compared against the corresponding temporally averaged reference fields. Annotations indicate the area-weighted percentage of ocean grid points with high temporal agreement ($R^2 \geq 0.8$).